\documentclass[12pt]{article}
\usepackage{amsmath} 
\usepackage{pstricks}
\usepackage{longtable}
\usepackage[dvips]{graphicx}
\usepackage{color}
\usepackage[tableposition=below]{caption}
\input{epsf} 
\newcommand\as{\alpha_{\mathrm{S}}}

\def\beq{\begin{equation}} 
\def\eeq{\end{equation}} 
 
\def\nn{\nonumber}

\def\b0{b_0}

\begin{document} 

\begin{titlepage}
\renewcommand{\thefootnote}{\fnsymbol{footnote}}
\begin{flushright}
ZU-TH 08/12
\end{flushright}
\par \vspace{10mm}

\begin{center}
{\Large \bf
Higgs production at the LHC:\\[0.5cm]
updated cross sections at $\sqrt{s}=8$ TeV}
\end{center}
\par \vspace{2mm}
\begin{center}
{\bf Daniel de Florian}${}^{(a)}$\footnote{deflo@df.uba.ar} and
{\bf Massimiliano Grazzini}${}^{(b)}$\footnote{grazzini@physik.uzh.ch}\footnote{On leave of absence from INFN, Sezione di Firenze, Sesto Fiorentino, Florence, Italy.}\\

\vspace{5mm}

${}^{(a)}$Departamento de F\'\i sica, FCEYN, Universidad de Buenos Aires, \\
(1428) Pabell\'on 1, Ciudad Universitaria, Capital Federal, Argentina\\

${}^{(b)}$Institut f\"ur Theoretische Physik, Universit\"at Z\"ürich, CH-8057 Z\"urich, Switzerland\\

\vspace{5mm}

\end{center}

\par \vspace{2mm}
\begin{center} {\large \bf Abstract} \end{center}
\begin{quote}
\pretolerance 10000

We present predictions for the inclusive cross section for Higgs boson production by gluon--gluon fusion
in proton collisions at $\sqrt{s}=8$ TeV.
Our calculation is accurate up to next-to-next-to-leading order in QCD perturbation theory and
includes soft-gluon effects up to next-to-next-to-leading logarithmic accuracy
and two-loop electroweak corrections.
The dependence on heavy-quark masses is taken into account exactly up to next-to-leading order and next-to-leading logarithmic accuracy, and a treatment of the Higgs boson line-shape is provided according to the complex-pole scheme.

\end{quote}

\vspace*{\fill}
\begin{flushleft}
June 2012
\end{flushleft}
\end{titlepage}

\setcounter{footnote}{1}
\renewcommand{\thefootnote}{\fnsymbol{footnote}}

The data collected at the LHC in 2011 allowed the ATLAS \cite{atlas} and CMS \cite{cms} collaborations to impose strong constraints on the allowed mass of the Standard Model (SM) Higgs \cite{Englert:1964et} boson, by essentially excluding it in the mass range 
${\cal O}(130\, {\rm GeV})< m_H<  {\cal O}(600\, {\rm GeV})$. Both collaborations observed an excess of events around $m_H \sim 125$ GeV and
the update of the Tevatron \cite{tevatron} results with up to 10 fb$^{-1}$ integrated luminosity also points to an excess in the region $115-135$ GeV.

With the LHC running at 8 TeV and expecting to deliver an integrated luminosity of ${\cal O}(15 \, \rm{fb}^{-1})$ per experiment, the discovery or exclusion of the SM Higgs boson can really be envisaged by the end of 2012.

The dominant mechanism for SM Higgs boson production at hadron colliders is gluon-gluon fusion \cite{Georgi:1977gs}, through a heavy-quark (mainly, top-quark) loop. The QCD radiative corrections to the total cross section have been computed at the next-to-leading order (NLO) in Refs.~\cite{Dawson:1990zj,Djouadi:1991tk,Spira:1995rr} and at the next-to-next-to-leading order (NNLO accuracy) in \cite{Harlander:2002wh,Anastasiou:2002yz,Ravindran:2003um}. NNLO results at the exclusive level can be found in Refs.~\cite{Anastasiou:2005qj,Anastasiou:2007mz,Catani:2007vq,Grazzini:2008tf}.

In this article we present state of the art predictions for this channel with explicit results at the LHC with centre-of-mass energy of 8 TeV.
%\footnote{Independent results at the same centre-of-mass energy, obtained at different levels of theoretical accuracy, are
%presented in Refs.~\cite{Anastasiou:2012hx,Baglio:2010ae}.}.
Other recent results for the Higgs production cross section at the LHC, obtained at different levels of theoretical accuracy, are presented in Refs.~\cite{Ahrens:2010rs,Baglio:2010ae,Anastasiou:2011pi,Anastasiou:2012hx}.
%presented in Refs.~\cite{Anastasiou:2012hx,Baglio:2010ae}.}.

The main features of our calculation have already been
described in Ref.~\cite{deFlorian:2009hc}.
Here we focus on the improvements with respect
to the work of Ref.~\cite{deFlorian:2009hc}.

Our calculation is based on the resummation of soft-gluon contributions to all orders, as a way to improve state of the art fixed-order predictions with the dominant effect from higher-order corrections.
The resummation of soft-gluon effects is achieved (see Ref.~\cite{Catani:2003zt} for more details) by organizing the partonic coefficient function in the $gg$ channel in Mellin space\footnote{The Mellin transform is defined with respect to the variable $z=m_H^2/{\hat s}$, ${\hat s}$ being the partonic centre-of-mass energy.} as
\begin{align}
\label{resfdelta}
G_{{gg},\, N}^{{\rm (res)}}(\as(\mu_R^2), m_H^2/\mu_R^2;m_H^2/\mu_F^2) 
&=  
\as^2(\mu^2_R)\,{C}_{gg}(\as(\mu^2_R),m_H^2/\mu^2_R;m_H^2/\mu_F^2) \nn \\ 
&\cdot  \Delta_{N}^{H}(\as(\mu^2_R),m_H^2/\mu^2_R;m_H^2/\mu_F^2) +
{\cal O}(1/N)\; , 
\end{align}
where $\as(\mu_R)$ is the QCD coupling evaluated at the renormalization scale $\mu_R$ and $\mu_F$ is the factorization scale.
The large logarithmic corrections (that appear as $\as^n\ln^{2n-k} N$ in Mellin space) are exponentiated in the  Sudakov radiative factor $\Delta_N^{H}$, which depends only on the dynamics of soft gluon emission from the initial state partons.
The hard coefficient ${C}_{gg}$ includes $N-$independent terms arising from both soft and hard gluon emission and depends on the details of the coupling to the Higgs boson  and, therefore, on the masses of the heavy quarks in the loop.
The coefficient needed to perform the calculation up to N$^i$LL can be obtained  from the corresponding fixed order computation to N$^i$LO accuracy.

The explicit expression for the coefficient ${C}_{gg}$ computed in the large-$m_t$ limit up to two-loop accuracy can be found in Ref.~\cite{Catani:2003zt}.
Together with the knowledge of the radiative factor $\Delta_N^{H}$, this result allowed us to perform the resummation up to next-to-next-to-leading logarithmic (NNLL) accuracy and to match the ensuing result with the fixed-order NNLO computation \cite{Harlander:2002wh,Anastasiou:2002yz,Ravindran:2003um} still performed in the large-$m_t$ limit \cite{Catani:2003zt}. The resummation effect has been confirmed by the computation of the soft-gluon terms at N$^3$LO \cite{Moch:2005ky}.
The NNLL+NNLO result of Ref.~\cite{Catani:2003zt} has been the reference theoretical prediction for the gluon fusion cross section for few years.

In Ref.~\cite{deFlorian:2009hc} we have extended this calculation by including the bottom-quark loop (and its interference with the top contribution) exactly up to NLO \cite{Djouadi:1991tk,Spira:1995rr}.
However, the NLO results in \cite{Spira:1995rr,Harlander:2005rq,Aglietti:2006tp,Anastasiou:2006hc} can be used to extract the {\em exact} expression of $C_{gg}$ in Eq.~(\ref{resfdelta}) up to NLL:
\begin{align}
\label{exactc1}
{C}_{gg}&\left(\as(\mu^2_R),m_H^2/\mu^2_R;m_H^2/\mu_F^2\right)=\nn\\
&~~~~~1 + \frac{\as(\mu_R^2)}{\pi} \left( c_{\phi}(m_q) + 6\zeta(2) +
\frac{33-2 N_f}{6}\ln\frac{\mu_R^2}{\mu_F^2} +
6 \gamma_E^2 + \pi^2 -6 \gamma_E \ln\frac{m_H^2}{\mu_F^2}\right)+{\cal O}(\as^2)
\end{align}
where $\gamma_E \simeq 0.577216$ is the Euler's constant and
the dependence on the heavy-quark masses appears in the function  $c_{\phi}(m_q)$. The corresponding expression can be found in Eq.~(B.2) of Ref.~\cite{Spira:1995rr} in terms of one-dimensional integrals, or as a fully analytic result in Eq.~(3.5) of Ref.~\cite{Harlander:2005rq} and Eq.~(27) of Ref.~\cite{Aglietti:2006tp}, both in terms of harmonic polylogarithms. In the limit of infinite quark mass, one recovers the well know result 
\begin{align}
 c_{\phi}(m_q) \xrightarrow[m_q\rightarrow \infty]{} \frac{11}{2} \, .
\end{align}
By using Eq.~(\ref{exactc1}) we can achieve NLL+NLO accuracy without relying on the large-$m_t$ approximation, that is, we can improve the exact fixed order NLO result by including soft-gluon resummation at NLL.
Since no exact results are available at NNLO accuracy,
at this order we only include the top-quark contribution in the $m_t \rightarrow \infty$ limit by adding soft-gluon effects at NNLL,
as in Refs.~\cite{Catani:2003zt,deFlorian:2009hc}.
The ensuing result is finally corrected for two-loop electro-weak (EW) contributions \cite{ew} as evaluated in \cite{Actis:2008ug}, in the {\em complete factorization} scheme,
in which the EW corrections are applied to the full QCD corrected cross section\footnote{Results including this improvement for $\sqrt{s}=7$ and $14$ TeV have been presented in Ref.~\cite{Dittmaier:2011ti}.}.
We point out that the inclusion of the exact dependence on the top- and bottom-quark masses up to NLL accuracy results in a decrease of the cross section
ranging from about $1.5\%$ at $m_H=125$ GeV,
%reaching up to $6.2\%$ at $m_H=1000$ GeV.
to about $6\%$ at $m_H=800$ GeV.
%
%%%%%% ADDED
%
The usually neglected charm-quark contribution, which we also include for the
first time in our calculation, further decreases the cross section by about $1\%$ for a light Higgs, being instead very small in the high-mass region.
%
%%%%%%%%%%%%%

The second improvement with respect to the work of Ref.~\cite{deFlorian:2009hc} regards the treatment
of the Higgs boson width.
While the Zero Width Approximation (ZWA) can be considered sufficiently accurate for the evaluation of the {\it inclusive} cross section for a light Higgs boson, the increase of the Higgs boson width at large masses requires a proper inplementation of the corresponding line-shape. In this work we rely on the OFFP scheme described in Ref.~\cite{offp} as an effective implementation of the complex-pole scheme.
The {\it signal} cross section can be written as
\begin{align}
\label{eq:lineshape}
\sigma(m_H)= \int dM^2 \frac{M \,\Gamma_H(M)}{\pi} \frac{{\tilde{\sigma}(M)}}{(M^2-m_H^2)^2+m_H^2 \gamma_H^2}\, ,
\end{align}
where $m_H+i \gamma_H$ parametrizes the complex pole of an unstable particle, with $m_H$ an input parameter  playing the role of the on-shell mass while $\gamma_H$, equivalent to the on-shell width, is computed at one loop accuracy in the SM in Ref.~\cite{offp}. $\Gamma_H(M)$ and ${\tilde{\sigma}(M)}$ correspond to the conventional on-shell width and hadronic production cross section evaluated at the virtuality of the Higgs boson $M$, respectively. The calculation in Ref.~\cite{offp} provides a realistic estimate of the complex-pole width $\gamma_H$ above the $ZZ$ threshold but might introduce an artificial effect at low masses due to the fact that in general $\Gamma(m_H)/\gamma_H \neq 1$. In order to recover the ZWA for light Higgs we use an extrapolation of the value of $\gamma_H$ towards the on-shell decay width $\Gamma(m_H)$ below $m_H=200$ GeV \cite{Passarino-private}
\footnote{Notice that effectively the OFFP scheme matches the naive Breit Wigner implementation below 200 GeV.}.

\small
%\begin{center}
\begin {longtable}{|c|c|c|c|}
\hline
$m_H$ (GeV) & $\sigma$ (pb) & scale($\%$) & PDF+$\as$($\%$)
\\
\hline
\endhead
%\\
115.0&22.68&+7.4 \, -8.1 &+7.6 \, -6.8\\ 
116.0&22.31&+7.4 \, -8.0&+7.5 \, -6.8\\ 
117.0&21.93&+7.4 \, -8.0&+7.5 \, -6.8\\ 
118.0&21.58&+7.4 \, -8.0&+7.5 \, -6.8\\ 
119.0&21.23&+7.3 \, -8.0&+7.5 \, -6.8\\ 
120.0&20.88&+7.3 \, -7.9&+7.5 \, -6.9\\ 
120.5&20.72&+7.3 \, -7.9&+7.5 \, -6.9\\ 
121.0&20.56&+7.3 \, -7.9&+7.5 \, -6.9\\ 
121.5&20.39&+7.3 \, -7.9&+7.5 \, -6.9\\ 
122.0&20.24&+7.3 \, -7.9&+7.5 \, -6.9\\ 
122.5&20.08&+7.2 \, -7.9&+7.5 \, -6.9\\ 
123.0&19.92&+7.2 \, -7.9&+7.5 \, -6.9\\ 
123.5&19.76&+7.2 \, -7.9&+7.5 \, -6.9\\ 
124.0&19.61&+7.2 \, -7.9&+7.5 \, -6.9\\ 
124.5&19.46&+7.2 \, -7.9&+7.5 \, -6.9\\ 
125.0&19.31&+7.2 \, -7.8&+7.5 \, -6.9\\ 
125.5&19.15&+7.2 \, -7.8&+7.5 \, -6.9\\ 
126.0&19.01&+7.2 \, -7.8&+7.5 \, -6.9\\ 
126.5&18.86&+7.2 \, -7.8&+7.5 \, -6.9\\ 
127.0&18.71&+7.1 \, -7.8&+7.5 \, -6.9\\ 
127.5&18.57&+7.1 \, -7.8&+7.5 \, -6.9\\ 
128.0&18.43&+7.1 \, -7.8&+7.5 \, -6.9\\ 
128.5&18.29&+7.1 \, -7.8&+7.5 \, -6.9\\ 
129.0&18.15&+7.1 \, -7.8&+7.5 \, -6.9\\ 
129.5&18.01&+7.1 \, -7.8&+7.5 \, -6.9\\ 
130.0&17.88&+7.1 \, -7.7&+7.5 \, -6.9\\ 
131.0&17.62&+7.1 \, -7.7&+7.5 \, -7.0\\ 
132.0&17.36&+7.0 \, -7.7&+7.5 \, -7.0\\ 
133.0&17.11&+7.0 \, -7.7&+7.4 \, -7.0\\ 
134.0&16.86&+7.0 \, -7.7&+7.4 \, -7.0\\ 
135.0&16.62&+7.0 \, -7.7&+7.4 \, -7.0\\ 
136.0&16.38&+6.9 \, -7.6&+7.4 \, -7.0\\ 
137.0&16.14&+6.9 \, -7.6&+7.4 \, -7.0\\ 
138.0&15.92&+6.9 \, -7.6&+7.4 \, -6.9\\ 
139.0&15.69&+6.9 \, -7.6&+7.4 \, -6.9\\ 
140.0&15.48&+6.9 \, -7.6&+7.4 \, -6.9\\                                                                              
150.0&13.53&+6.7 \, -7.4&+7.4 \, -7.0\\
160.0&11.85&+6.5 \, -7.3&+7.5 \, -7.1\\
180.0&8.810&+6.2 \, -7.0&+7.4 \, -7.5\\                                                                       
190.0&7.815&+6.1 \, -6.9&+7.4 \, -7.5\\                                       
200.0&7.082&+6.0 \, -6.8&+7.4 \, -7.7\\                                      
210.0&6.496&+6.0 \, -6.7&+7.4 \, -7.8\\                                                                              
220.0&6.005&+5.9 \, -6.6&+7.3 \, -7.6\\
\hline                                                                              
230.0&5.565&+5.9 \, -6.5&+7.4 \, -7.7\\                                                                              
240.0&5.158&+5.9 \, -6.4&+7.3 \, -7.7\\                                                                              
250.0&4.781&+5.8 \, -6.4&+7.4 \, -7.7\\
260.0&4.460&+5.8 \, -6.3&+7.6 \, -7.4\\
270.0&4.182&+5.8 \, -6.2&+7.6 \, -7.9\\                                                                              
280.0&3.950&+5.7 \, -6.2&+7.6 \, -8.0\\                              
290.0&3.754&+5.7 \, -6.1&+7.6 \, -8.0\\                                       
300.0&3.595&+5.7 \, -6.1&+7.7 \, -7.9\\
310.0&3.472&+5.7 \, -6.0&+7.7 \, -8.0\\                                                                              
320.0&3.383&+5.7 \, -6.0&+7.7 \, -8.0\\
330.0&3.341&+5.7 \, -6.0&+7.8 \, -8.1\\                                                                              
340.0&3.359&+5.7 \, -5.9&+7.9 \, -8.1\\                                                                              
350.0&3.399&+5.7 \, -5.9&+8.0 \, -8.2\\                                       
360.0&3.384&+5.8 \, -5.9&+8.0 \, -8.2\\
370.0&3.331&+5.8 \, -5.8&+8.1 \, -8.2\\
380.0&3.231&+5.8 \, -5.6&+8.1 \, -8.2\\  
390.0&3.089&+5.8 \, -5.5&+8.2 \, -8.2\\    
400.0&2.921&+5.8 \, -5.4&+8.2 \, -8.2\\  
420.0&2.550&+5.8 \, -5.3&+8.3 \, -8.3\\    
440.0&2.179&+5.8 \, -5.3&+8.5 \, -8.4\\      
450.0&2.002&+5.8 \, -5.2&+8.6 \, -8.4\\
460.0&1.836&+5.8 \, -5.2&+8.7 \, -8.4\\  
480.0&1.537&+5.8 \, -5.2&+8.9 \, -8.5\\     
500.0&1.283&+5.8 \, -5.1&+9.1 \, -8.5\\ 
520.0&1.069&+5.8 \, -5.1&+9.2 \, -8.6\\ 
540.0&0.8911&+5.8 \, -5.1&+9.4 \, -8.6\\       
550.0&0.8141&+5.8 \, -5.1&+9.4 \, -8.7\\     
560.0&0.7442&+5.9 \, -5.1&+9.4 \, -8.7\\   
580.0&0.6230&+5.9 \, -5.1&+9.5 \, -8.7\\  
600.0&0.5231&+5.9 \, -5.0&+9.5 \, -8.8\\      
620.0&0.4403&+5.9 \, -5.0&+9.6 \, -8.9\\    
640.0&0.3719&+5.9 \, -5.0&+9.7 \, -9.0\\     
650.0&0.3424&+5.9 \, -5.0&+9.7 \, -9.0\\ 
660.0&0.3153&+5.9 \, -5.1&+9.8 \, -9.1\\  
680.0&0.2680&+6.0 \, -5.1&+9.9 \, -9.2\\  
700.0&0.2289&+6.0 \, -5.1&+10.1 \, -9.3\\  
720.0&0.1962&+6.0 \, -5.1&+10.2 \, -9.5\\   
740.0&0.1687&+6.1 \, -5.1&+10.4 \, -9.6\\ 
750.0&0.1566&+6.1 \, -5.1&+10.4 \, -9.7\\  
760.0&0.1455&+6.1 \, -5.2&+10.5 \, -9.7\\     
780.0&0.1260&+6.1 \, -5.2&+10.5 \, -9.8\\      
800.0&0.1095&+6.1 \, -5.2&+10.6 \, -9.8\\
\hline
%\captionsetup{width=16cm}
\caption{Cross sections at the LHC ($\sqrt{s}=8$ TeV) and corresponding scale and PDF+$\as$ uncertainties computed according to the PDF4LHC recommendation.}
\end{longtable}
\normalsize

The results we are going to present are obtained by using the MSTW2008 NNLO parton distribution functions (PDFs) \cite{Martin:2009iq},
setting the
reference values for the factorization and renormalization scales to the Higgs boson virtuality $M$ \footnote{The numerical integration over the virtuality $M$ in Eq.~(\ref{eq:lineshape}) is performed between 50 to 1800 GeV.}.
The on-shell width of the Higgs boson is evaluated with the program HDECAY \cite{Djouadi:1997yw}.
We set the top-quark mass to $m_t=172.5$ GeV and we choose $m_b=4.75$ GeV and $m_c=1.40$ GeV consistently with the MSTW2008 set.

Our predictions and the corresponding uncertainties, computed as discussed below, are presented in Table 1.
We stress that the inclusion of finite-width effects results in an increase of the cross section with respect to the ZWA of about ${\cal O}(10\%)$ at $m_H=800$ GeV.
It is well known that as $m_H$ increases and finite-width effects become important, the signal cross section
becomes itself ill defined,
and for each decay channel of the Higgs boson, only the full signal + background computation in that
channel strictly makes sense\footnote{A method to estimate the uncertainty from interference effects in the $ZZ$ channel is proposed in Ref.~\cite{giampiero}.}.
The use of a naive Breit Wigner, which would correspond to replace $\gamma_H$ with the on-shell width $\Gamma(m_H)$ in Eq.~(\ref{eq:lineshape}), would give a smaller cross section with respect to the result in the complex-pole scheme, the difference ranging from $-3.5\%$ for $m_H=300$ GeV to $-18\%$ at $m_H=600$ GeV, to $-27\%$ at $m_H=800$ GeV. 

%%%% Uncertainties

We now review the various sources of uncertainty affecting the 
cross sections presented in Table 1. The uncertainty has two main origins:
the one coming from the partonic cross sections, and the one arising from our limited knowledge of the PDFs.

Uncalculated higher-order QCD radiative corrections are the most important source of uncertainty
on the partonic cross section. A method, which is customarily used in perturbative QCD
calculations, to estimate their size is to vary the renormalization and factorization scales around
the hard scale $M$ \footnote{An attempt to go beyond this standard approach is made in Ref.~\cite{Cacciari:2011ze}.}.
In general, this procedure can only give a lower limit on the {\it true}
uncertainty.
Here we quantify the uncertainty as in Refs.~\cite{Catani:2003zt,deFlorian:2009hc}: we vary
independently $\mu_F$ and $\mu_R$ 
in the range $0.5 M\leq \mu_F,\mu_R\leq 2 M$, with the constraint
$0.5 \leq \mu_F/\mu_R \leq 2$. 

The scale uncertainty ranges from $+7-8\%$ ($m_H=125$ GeV)
to about $+6-5\%$ ($m_H=800$ GeV).
The results are
consistent with those of Ref.~\cite{Catani:2003zt,deFlorian:2009hc};
in particular, we note that the effect of scale variations
in our resummed calculation is considerably
reduced with respect to the corresponding fixed-order NNLO result.

Another source of perturbative uncertainty on the partonic cross sections comes from the implementation of
the EW corrections. Our results are obtained in the complete factorization scheme, a scenario supported by the effective field theory computation of mixed QCD-EW corrections presented in Ref.~\cite{Anastasiou:2008tj}.
The partial factorization scheme, in which EW corrections are applied only to the LO cross section, would lead to a change of our results varying
from about $-3 \%$ to $+2\%$ in the range of Higgs boson masses we consider.

A potentially important
source of perturbative uncertainty arises from the use of the
large-$m_t$ approximation in the computation of the partonic cross section beyond NLL+NLO.
The accuracy of the large-$m_t$ approximation at NNLO
has been studied by computing subleading terms in the large-$m_t$ limit \cite{largemtop}, concluding that it works remarkably well, to better than $1\%$ for $m_H<300$ GeV.
For heavier Higgs bosons, we expect the uncertainty due to the large-$m_t$ approximation to play an increasing role.
Nonetheless, it is well known that the comparison of the exact NLO calculation with the result obtained in the large $m_t$ limit but rescaled with the exact ($m_t$ and $m_b$ dependent) LO cross section shows agreement within the $10\%$ level even at high Higgs boson masses \cite{Kramer:1996iq}.
Since our calculation exactly includes the heavy-quark mass dependence up to NLL+NLO, the uncertainty due to the large-$m_t$ approximation should be well below the $10\%$ effect in the mass range we
consider.

The other important source of uncertainty in the cross section is
the one coming from PDFs. Our understanding of PDFs has improved considerably
in the last years, and we have now various PDF sets at NNLO accuracy:
MSTW2008 \cite{Martin:2009iq}, NNPDF21 \cite{Ball:2011uy}, JR09 \cite{JimenezDelgado:2009tv} and ABK11 \cite{Alekhin:2012ig}.
In order to produce the central values for the cross section, we rely on the 
MSTW2008 NNLO PDFs \cite{Martin:2009iq}. The PDF and $\as$ uncertainties are estimated using the corresponding $68\%$ C.L. band from the MSTW sets, normalized according to the PDF4LHC prescription \cite{Botje:2011sn}. We note that the ensuing uncertainty is rather close to the one obtained by using the $90\%$ C.L. set of MSTW. The uncertainty ranges from $+8-7\%$ ($m_H=125$ GeV)
to $+11-10\%$ ($m_H=800$ GeV).
By using the NNPDF21 NNLO default set \cite{Ball:2011uy} with $\as(m_Z)=0.119$ we find differences that range from $+5\%$ ($m_H=125$ GeV) to $+2\%$ ($m_H=800$ GeV) with respect to our central MSTW2008 result.
When using the NNPDF21 NNLO set with $\as(m_Z)=0.114$ the difference
ranges from $-3\%$ ($m_H=125$ GeV) to $-11\%$ ($m_H=800$ GeV).

The JR09 and ABM11 partons give larger differences with respect to our central MSTW prediction. For $m_H=125$ GeV the JR09 (ABM11) result is lower than MSTW08 by about $10\%$ ($7\%$). At larger Higgs masses the agreement of JR09 with MSTW08 improves, but the difference with ABM11 increases, being about $14\%$ at $m_H=300$ GeV and further increasing at higher Higgs masses.
We remind the reader that the ABM11 NNLO fit does not include Tevatron jet data and that the ensuing QCD coupling is $\as(m_Z)=0.1134$,
%quite below from the world average.
significantly smaller than the world average. 

Recently also the CTEQ collaboration has released an NNLO PDF set \cite{cteq}.
We find that the use of the CT10 NNLO central set, corresponding to $\as(m_Z)=0.118$ leads to results that agree at the $1\%$ level or better with those obtained with MSTW2008.
By using the sets corresponding to $\as=0.116$ and $\as=0.120$ the differences range from $\pm 4\%$ at $m_H=125$ GeV to $+5-6\%$ at $m_H=800$ GeV, thus well within our PDF4LHC uncertainty band. We find this agreement
reassuring and we conclude that
our central predictions, endowed with the PDF4LHC uncertainty, should be sufficiently conservative. Nonetheless, we believe that the large differences obtained with JR09 and ABM11 PDFs definitely deserve further investigations.

%\begin{figure}[!ht]
%\centering
%\includegraphics[width=0.5\textwidth]{h.ps}
%\caption{{\em gluon-gluon fusion Higgs cross section at the LHC with 8 TeV. The inner band (blue) represents the TH uncertainty  while the outer one (red) corresponds to the total uncertainty}}
%\label{fig:H}
%\end{figure}

The results of this paper can be compared to those presented in 
Ref.~\cite{Anastasiou:2012hx}, where the impact of finite-width effects is studied as well.
Besides the NLO QCD corrections with the exact dependence on the top- and bottom-quark masses, the NNLO corrections in the large-$m_t$ limit, and two loop EW effects \cite{Actis:2008ug}, the calculation of Ref.~\cite{Anastasiou:2012hx} includes
mixed QCD-EW corrections evaluated in an effective field theory approach \cite{Anastasiou:2008tj}
and the independent evaluation of EW effects from real radiation \cite{Keung:2009bs,Brein:2010xj} whose effect is, however, at the $1\%$ level or smaller, and that we neglect here. For a light Higgs boson, the main difference with our computation arises from the evaluation of higher-order QCD corrections.
Following what previously done in Ref.~\cite{Anastasiou:2008tj}, in Ref.~\cite{Anastasiou:2012hx} these corrections are computed up to NNLO
but choosing $\mu_F=\mu_R=m_H/2$,
as an attempt to reproduce effects beyond NNLO, that, in our calculation, are instead estimated through soft-gluon resummation.
For $m_H=125$ GeV the result of Ref.~\cite{Anastasiou:2012hx} with the corresponding scale uncertainty is $\sigma=20.69^{+8.4\%}_{-9.3\%}$ pb, $7\%$ higher with respect to our $\sigma=19.31^{+7.2\%}_{-7.8\%}$ pb, but still well within the uncertainty bands.
Larger differences are observed in the high mass region, due to the different implementation of finite-width effects. In Ref.~\cite{Anastasiou:2012hx} a Breit-Wigner with running width is used
as the default implementation of the line-shape. At $m_H=400$ GeV, the result 
of Ref.~\cite{Anastasiou:2012hx} is about $16\%$ smaller than ours.

In this paper we have presented updated predictions for the cross section
for Higgs boson production at the LHC with $\sqrt{s}=8$ TeV, and discussed the corresponding uncertainties. The results are based
on the most advanced theoretical information available at present for this observable, including 
soft-gluon resummation up to NNLL accuracy, two-loop EW corrections, exact treatment of heavy quark mass effects up to NLL+NLO accuracy and finite-width effects evaluated in the complex-pole scheme.
We look forward to a comparison of our results with LHC data.

We would like to thank Giampiero Passarino and Carlo Oleari for useful discussions. This work was supported in part by UBACYT, CONICET, ANPCyT and the Research Executive Agency (REA) of the European Union under the Grant Agreement number PITN-GA-2010-264564 (LHCPhenoNet).

\end{document}